\documentclass[12pt,fleqn]{article}

\usepackage{epsfig}
\usepackage{graphics}
\usepackage{latexsym}
\usepackage{amssymb}
\usepackage{amsmath}
\def\i{\mathrm{i}}
\def\e{\mathrm{e}}
\def\Qm{Q_\mathrm{min}}
\def\Qmc{\overline{Q}_\mathrm{min}}
\def\Hm{H_\mathrm{min}}
\def\Hmc{\overline{H}_\mathrm{min}}
\def\slim{\mathop{\mathrm{s-}}\!\lim}
\newcommand{\ben}{\begin{displaymath}}
\newcommand{\een}{\end{displaymath}}
\newtheorem{lemma}{Lemma}
\newtheorem{theorem}[lemma]{Theorem}

\newtheorem{proposition}[lemma]{Proposition}
\title{No zero energy states for the supersymmetric $x^2y^2$ potential}
\author{ G.M. Graf${}^{(a)}$,  D. Hasler${}^{(a)}$,  J.
Hoppe${}^{(b)}$ \\ 
\vspace*{-0.05truein} \\
\normalsize\it ${}^{(a)}$ Theoretische Physik,
ETH-H\"onggerberg, CH--8093 Z\"urich\\
\normalsize\it   ${}^{(b)}$ QFT, HU-Berlin, Invalidenstrasse 110, 
D--10115 Berlin}

\begin{document}

\maketitle
\vspace{0.4cm}
\begin{abstract}
We show that the positive supersymmetric matrix-valued differential
operator
$H={p_x}^2 + {p_y}^2 + x^2y^2 + x \sigma_3 + y \sigma_1$ has no zero modes, 
i.e., $H \psi = 0$ implies $\psi =0$.
\end{abstract}

\section{Introduction}
\label{sec:intr}
The Hamiltonian of the model is plainly given as   
\ben
H={p_x}^2 + {p_y}^2 + x^2y^2 + x \sigma_3 + y \sigma_1 \ ,
\een
acting on the Hilbert space 
$\mathcal H ={\mathrm L}^2(\mathbb R^2) \otimes \mathbb
C^2$, where
$\sigma_i$ are the Pauli matrices. The supercharge is
\begin{equation}
Q = p_x \sigma_3 - p_y \sigma_1 - xy \sigma_2\ .
\label{q}
\end{equation} 
Together with the reflection
\ben
( P \psi)(x,y) =  \frac{1}{\sqrt{2}}(\sigma_1 +\sigma_3) \psi (y,x) \ , 
\een
the system ($H,P,Q$) exhibits
supersymmetry:  
\ben
H=Q^2\ ,\qquad P^2 = 1\ ,\qquad  QP + PQ = 0  \ .
\een
It was shown in \cite{dewitetal}\footnote{The Pauli matrices used there 
differ from ours by an irrelevant unitary conjugation.} that the spectrum of 
$H$ is $\sigma(H)=[0,\infty)$. The question whether $0$ is an eigenvalue of
$H$ has however so far eluded a definite answer --- despite some efforts 
in this direction such as \cite{ak}.

The precise definition of the model is as follows: $Q$
is the closure of (\ref{q}) on $C_0^{\infty}\otimes \mathbb
C^2$, where $\mathcal{C}_0^{\infty} \equiv
\mathcal{C}_0^{\infty}(\mathbb{R}^2)$, and $H$ is the self-adjoint
operator
associated to the positive symmetric quadratic form  
$q(\varphi,\psi) = (Q \varphi, Q \psi)$ on $\mathcal{D}(Q) \times
\mathcal{D}(Q)$. Hence $C_0^{\infty}\otimes \mathbb C^2$ is a form core
for $q$. The main result is the following.

\begin{theorem}
\label{nozm}
Let $\psi \in \mathcal{D}(H)$ with $H \psi = 0$. Then $\psi = 0$.
\end{theorem}

The proof relies on a simple commutator argument. Let $f =
f(x,y)$ be real-valued. Since $H \psi = 0$ implies $Q \psi = 0 $ we then have
\begin{eqnarray*}
( \psi, i[Q,f] \psi) & = & \i\{ (Q \psi,f\psi) - (f \psi , Q \psi )
\} = 0 \ ,\\ 
\i [Q,f] & = & \sigma_3 \partial_x f - \sigma_1 \partial_y f
\end{eqnarray*}
for `any' function $f$. In particular for $f = (x^2 - y^2)/2$
we get
\ben
(\psi, (x \sigma_3 + y \sigma_1 ) \psi ) = 0 \ .
\een
Subtracting this from $(\psi,H \psi) = 0$ we obtain 
\ben
(\psi, (p_x^2 + p_y^2 + x^2 y^2) \psi ) = 0 \ ,
\een
which is impossible unless $\psi = 0$, since $p_x^2 + p_y^2 + x^2 y^2 >
0$ \cite{bsimon1}. The trouble with this is that $\psi$ may fail to be 
in the domain of $f = (x^2 - y^2)/2$ or, what matters more, in
that of
$\i[Q,f] = x \sigma_3 + y \sigma_1$. The proof given in the next section will
circumvent this difficulty.

We stress that any natural realization of $Q$ and $H$ coincides with the 
one given above. This is the content of the following statement, which is 
however not needed for the main result. 

\begin{proposition}
\label{char} \
\begin{description}  
\item[(a)] Q is self-adjoint, and its domain is 
\ben 
\mathcal{D}(Q) = \{ \psi \in \mathcal{H}
\ | \ Q \psi \in \mathcal{H} \textrm{ (in the sense of distributions on
$C_0^{\infty}\otimes \mathbb C^2$)}  \} \ .
\een 
\item[(b)] $\mathcal{D}(H) =  \{ \psi \in \mathcal{H}
\ | \ H \psi \in \mathcal{H} \textrm{ (in the sense of distributions on
$C_0^{\infty}\otimes \mathbb C^2$)}  \}$.
\item[(c)] $C_0^{\infty}\otimes \mathbb C^2$ is an operator core for
$H$.
\item[(d)] $H=Q^2$, i.e., $\psi \in \mathcal{D}(H)$ iff $\psi \in
\mathcal{D}(Q)$ and $Q \psi \in \mathcal{D}(Q)$, in which case 
$H \psi =Q^2\psi$.
\end{description}  
\end{proposition}

Before turning to the proofs we give a simple argument showing
that a possible zero mode of $H$ cannot be unique, though this statement is 
superseded by Theorem \ref{nozm}. The operators on $\mathcal{H}$
\ben
(P_1\psi)(x,y)=\sigma_1\psi(-x,y)\ ,\;
(P_2\psi)(x,y)=\sigma_2\psi(-x,-y)\ ,\;
(P_3\psi)(x,y)=\sigma_3\psi(x,-y)
\een
satisfy $P_i^2=1,\,[P_i,P_j]=2\i\varepsilon_{ijk}P_k$ and $[Q,P_i]=0$. Thus 
$Q\psi=0$ implies $QP_i\psi=0$ and, if uniqueness is assumed, 
$P_i\psi=s_i\psi$ with $s_i=\pm 1$. But this contradicts the commutation
relations of the $P_i$.

A related argument shows that the index $\textrm{tr}( P \Pi )$, where $\Pi$ is
the ground state projection of $H$, vanishes. This is seen from $P P_2 = - P_2
P$ 
and 
\ben
\mathrm{tr} ( P \Pi ) = \mathrm{tr} ( P \Pi P_2^2 ) = - \mathrm{tr} ( P \Pi
P_2^2 ) \ ,
\een 
where one power of $P_2$ has been turned around the trace.

Finally we remark that absence of zero energy states had been suggested by the 
asymptotic analysis of \cite{fghhy} and, for a slightly less elementary model,
been proven in \cite{fh} by different means. 

\section{Proofs}
\label{sec:pr}
{\bf Proof of Theorem \ref{nozm}.}
Let 
\begin{eqnarray*}
h(x) = \left\{ \begin{array}{ll}
			-Mx -\frac1 2 M^2       & \ \  \ (x \le -M) \\
			\frac1 2 x^2 		& \ \ \ (-M \le x \le M) \\ 
			Mx - \frac1 2 M^2	& \ \ \ (x \ge M) 
		\end{array} \right.  \ ,
\end{eqnarray*}
whence
\begin{eqnarray*}
{h}'(x) & = &
 \left\{ \begin{array}{ll}
			-M              & \ \ \  (x \le -M) \\
			x 		&  \ \ \ (-M \le x \le M) \\
			M	        & \ \ \  (x \ge M)
		\end{array} \right. \\
        & = & x - g(x) \ ,
\end{eqnarray*}
where we have defined $g$ as
\begin{eqnarray*}
g(x) = \left\{ \begin{array}{ll}
			x + M & \ \ \ (x \le -M) \\
			0 		& \ \ \ (-M \le x \le M) \\
			x - M 	& \ \ \ (x \ge M)
		\end{array} \right. \ .
\end{eqnarray*}
Let furthermore $h_{\epsilon}(x) = h(x)\e^{-\epsilon \sqrt{1 + x^2}}, \, 
f_{\epsilon}(x,y)  = h_{\epsilon}(x) - h_{\epsilon}(y)$.
For $\varphi, \psi \in \mathcal{D}(Q)$ we have
\begin{equation}
\label{qf-fq}
\i[ ( Q\varphi,f_{\epsilon} \psi) - ( f_{\epsilon} \varphi, Q \psi ) ] =
(\varphi,( \sigma_3 \partial_x f_{\epsilon} - \sigma_1 \partial_y
f_{\epsilon} ) \psi ) \ .
\end{equation}
This equality is straightforward for $\varphi,\psi \in
\mathcal{C}_0^{\infty}\otimes \mathbb{C}^2$ and extends to
$\mathcal{D}(Q)$,
since the operator on the right side is bounded and
$\mathcal{C}_0^{\infty}\otimes \mathbb{C}^2$
is an operator core for $Q$.
By dominated convergence, 
\begin{equation}
\label{slim}
\slim_{\epsilon \downarrow 0} 
( \sigma_3 \partial_x f_{\epsilon} - \sigma_1 \partial_y
f_{\epsilon} ) = {h}'(x) \sigma_3 + {h}'(y) \sigma_1 \ .
\end{equation}
Subtracting the r.h.s. from $H$, we obtain 
\begin{eqnarray}
 H -  {h}'(x) \sigma_3 - {h}'(y) \sigma_1  
& = & p_x^2+ p_y^2 + x^2y^2 + g(x)\sigma_3 + g(y) \sigma_1 \nonumber \\
& \ge & p_x^2 + p_y^2 + x^2y^2 - |g(x)| - |g(y)| \equiv H_M \ .
\label{lb}
\end{eqnarray}
The inequality is understood in the sense of forms \cite{rs4,k},
where
$H_M$ is the self-adjoint operator associated to the corresponding form
with
form core  $\mathcal{C}_0^{\infty}\otimes \mathbb{C}^2$.
We claim that for $M$ large enough $H_M$ is positive,
i.e., $H_M > 0$. Now let $\psi \in \mathcal{D}(H)$ with $H \psi = 0$. This
implies $Q \psi = 0$ and, by (\ref{qf-fq}) and (\ref{slim}), 
\ben
(\psi, ( H -  {h}'(x) \sigma_3 - {h}'(y) \sigma_1 ) \psi ) = 0 \ .
\een
By  (\ref{lb}) and the claim this is possible only if $\psi = 0$ and the
theorem follows.
To prove the claim, we consider the partition of $\mathbb{R}^2$ as in
the figure 
\begin{figure}[h]
\begin{center}  
        \input{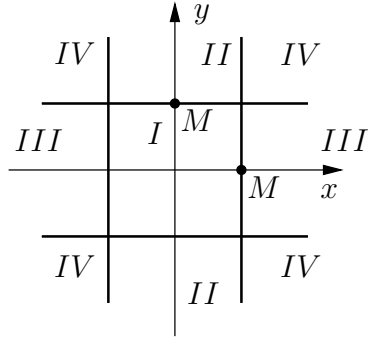}
        \caption{Partition of $\mathbb{R}^2$}
        \label{fig:GL1}
\end{center}
\end{figure}
and introduce Neumann conditions along the boundaries. Then (\cite{rs4},
XIII.15, Proposition 4)
\ben
H_M \ge H_I + H_{II} + H_{III} + H_{IV}  \ ,
\een
where the forms 
\begin{eqnarray*}
& H_{I} & =  p_x^2 + p_y^2 + x^2 y^2 \\
& H_{II} & =  p_x^2 + p_y^2 + x^2 y^2 - |y| + M \\
& H_{III} & =  p_x^2 + p_y^2 + x^2y^2 - |x| + M \\
& H_{IV} & =  p_x^2 + p_y^2 + x^2y^2 - |x| - |y| + 2M 
\end{eqnarray*}
act on the corresponding regions.
We show that $H_a > 0$ for $a = I,\ldots IV$ and $M$ large enough.

$I$. We have $H_I \ge 0$, and $(\psi, H_I \psi) = 0$ implies 
$(\psi, x^2 y^2 \psi)=0$ and hence $\psi = 0$.

$II$. The operator $p_x^2 + x^2$ on ${\mathrm L}^2(-a,a)$ with Neumann
boundary conditions at $x=\pm a$ satisfies
\ben
p_x^2 + x^2 \ge 1 - Ca^{-2} \ ,
\een
where $C$ denotes a generic constant. This can e.g. be seen by means of a
partition of unity $j_1^2 +j_2^2 =1$ with $j_1=j_1(x/a)$ equal to $1$ near 
$x/a=0$ and to $0$ near $x/a=\pm 1$. Hence, by scaling, $p_x^2 + x^2y^2$ on 
${\mathrm L}^2(-M,M)$ is estimated from below as
\ben
p_x^2 + x^2y^2  \ge  |y| ( 1 - C(|y|^{1/2}M)^{-2}) 
  =  |y| - C M^{-2}  \ .
\een
As a result, $H_{II} \ge M - CM^{-2}$, which is positive for $M$ large enough.

$III$. Is analogous to case $II$.

$IV$. There we have $x^2y^2 \ge M^3|x|,\,M^3|y|$ and hence
\ben
x^2y^2 - |x| - |y|  \ge  \Bigl( \frac1 2 M^3 - 1 \Bigr) (|x| + |y|) 
 \ge  \Bigl( \frac1 2 M^3 - 1 \Bigr) 2 M \ ,
\een
which is again positive for $M$ large enough.

\medskip \noindent
\noindent {\bf Proof of Proposition \ref{char}.} (a) Let 
$\mathcal{D}(\Qm) = \mathcal{C}_0^{\infty} \otimes
\mathbb{C}^2$, then $\Qmc = Q$ by definition of $Q$ and 
\[ \mathcal{D}(\Qm^*) =  \{ \psi \in \mathcal{H}
\ | \ Q \psi \in \mathcal{H} \ \textrm{(in the sense of distributions on 
$\mathcal{C}_0^{\infty}\otimes \mathbb{C}^2$)} \} \ . 
\]
Since $\Qm$ is symmetric, 
$\Qmc\subset \Qm^*$. We will show that 
\begin{equation}
\Qm^*\subset \Qmc \ . 
\label{QQ}
\end{equation}
This implies $\Qmc=\Qm^*=\Qmc^*$. It remains to prove (\ref{QQ}).
We pick $f \in \mathcal{C}_0^{\infty}$ with
$f(0) = 1$, such that $\slim_{n \to \infty}f_n = 1,\ \| \nabla f_n\|_{\infty} 
\to 0$ for $f_n(\vec{x}) = f(\vec{x}/n)$, and set 
$\tilde{f}_n (\vec{p})=f(\vec{p}/n^2)$. We approximate a given 
$\psi\in\mathcal{D}(\Qm)$ by 
$\psi_n= f_n \tilde{f}_nf_n\psi\in\mathcal{C}_0^{\infty}\otimes\mathbb{C}^2$. 
with $n\to\infty$. We have 
\ben
Q \psi_n = f_n\tilde{f}_n f_nQ \psi + [ Q,  f_n \tilde{f}_nf_n] \psi\ ,
\een
as one checks by taking inner products with 
$\varphi\in \mathcal{C}_0^{\infty}\otimes\mathbb{C}^2$. Here
\begin{eqnarray*}
[Q,f_n \tilde{f}_nf_n]&=&
[Q,f_n] \tilde{f}_nf_n + f_n [Q,\tilde{f}_n]f_n+f_n\tilde{f}_n[Q,f_n]\ , \\ 
{[Q,f_n] }
& = & -\i(\partial_x f_n) \sigma_3 +  \i (\partial_y f_n)\sigma_1\ , \\
{[Q,\tilde{f}_n]}& = & [xy,\tilde{f}_n] 
=\i\bigl( y(\partial_{p_x}\tilde{f}_n)+x(\partial_{p_y}\tilde{f}_n)\bigr)
+ \partial^2_{p_x p_y} \tilde{f}_n 
\end{eqnarray*}
are bounded operators with 
$\|[Q,f_n]\|\to 0$ and, due to 
$|x|,\,|y| \le C n$ on supp$f_n$, 
\ben
\| f_n [Q,\tilde{f}_n ] \| \le  C  n \cdot \frac{1}{n^2} + C
\frac{1}{n^4} \to 0 \ .
\een
Hence $\psi_n\to \psi$ and $Q\psi_n \to Q\psi$, i.e.,
$\psi
\in \mathcal{D}(\Qmc)$.

(d) By definition of the operator $H$ associated to the form $q$ we have:
$\psi\in \mathcal{D}(H)$ iff 
\begin{equation}
\psi\in \mathcal{D}(Q) \quad\textrm{and}\quad\exists \varphi  \in
\mathcal{H} \ \forall\eta\in \mathcal{D}(Q):\ (Q\psi,Q\eta)=(\varphi,\eta)\ ,
\label{cond}
\end{equation}
in which case $H\psi=\varphi$. This condition is also equivalent to 
$Q\psi\in \mathcal{D}(Q^*)$, with $Q^*Q\psi=\varphi$ in case of validity. That
proves (d). 

(b) In (\ref{cond}) one can replace $\mathcal{D}(Q)\ni\eta$ by the 
core $\mathcal{C}_0^{\infty} \otimes \mathbb{C}^2$, which proves (b) with
$\mathcal{H}$ replaced by $\mathcal{D}(Q)$. Left to show is that if 
$\psi \in \mathcal{H}$ with 
$H \psi\, (= Q^2 \psi)\in\mathcal{H}$ (in the sense of distributions), then 
$Q \psi \in\mathcal{H}$ (in the same sense). 
By elliptic regularity, $f_n^2 Q \psi\in\mathcal{D}(Q)$ and 
\ben
Qf_n^2 Q \psi=f_n^2Q^2\psi+2[Q,f_n]f_nQ\psi\ .
\een
This implies $\|f_n Q\psi\|^2\le C\|\psi\|(\|H\psi\|+\|f_nQ\psi\|)$ and hence
that $\|f_nQ\psi \|$ is bounded. We conclude $Q\psi\in\mathcal{H}$ by monotone
convergence.
 
(c) Let $\mathcal{D}(\Hm) = \mathcal{C}_0^{\infty} \otimes \mathbb{C}^2$. 
By (b), $\Hm^*=H$ and thus $\Hmc = \Hm^{**}= H^*=H$.

\medskip \noindent 
\noindent {\bf Acknowledgments.\/} We thank J. Fr\"ohlich for useful 
discussions.

\end{document}